\documentstyle[prb,aps,preprint,tighten]{revtex}

\begin{document}
\preprint{cond-mat/9504047}
\draft
\title{Weak localization coexisting with a magnetic field \\ in a
normal-metal--superconductor microbridge}
\author{P. W. Brouwer and C. W. J. Beenakker}
\address{Instituut-Lorentz, University of Leiden, P.O. Box 9506, 2300 RA
Leiden, The Netherlands}
\maketitle%

\begin{abstract}
A random-matrix theory is presented which shows that breaking time-reversal
symmetry by itself does {\em not} suppress the weak-localization correction to
the conductance of a disordered metal wire attached to a superconductor.
Suppression of weak localization requires applying a magnetic field as well as
raising the voltage, to break both time-reversal symmetry and electron-hole
degeneracy. A magnetic-field dependent contact resistance obscured this anomaly
in previous numerical simulations.
\medskip
%
\end{abstract}


Weak localization is a quantum correction of order $e^2/h$ to the classical
conductance of a metal.\cite{Anderson,Gorkov} The word ``localization'' refers
to the negative sign of the correction,\cite{SO} while the adjective ``weak''
indicates its smallness. In a wire geometry the weak-localization correction
takes on the universal value\cite{MelloStone} $\delta G = -\frac{2}{3}e^2/h$ at
zero temperature, independent of the wire length $L$ or mean free path
$\ell$.\cite{WIRE} The classical (Drude) conductance $G_0 \simeq (N\ell/L)\,
e^2/h$ is much greater than $\delta G$ in the metallic regime, where the number
of scattering channels $N \gg L/\ell$. Theoretically, the weak-localization
correction is the term of order $N^{0}$ in an expansion of the average
conductance $\langle G \rangle = G_0 + \delta G + {\cal O}(N^{-1})$ in powers
of $N$. Experimentally, $\delta G$ is measured by application of a weak
magnetic field $B$, which suppresses the weak-localization correction but
leaves the classical conductance unaffected.\cite{Bergmann} The suppression
occurs because weak localization requires time-reversal symmetry (${\cal T}$).
In the absence of ${\cal T}$, quantum corrections to $G_0$ are of order
$N^{-1}$ and not of order $N^{0}$. As a consequence, the magnetoconductance has
a dip around $B=0$ of magnitude $\delta G$ and width of order $B_c$ (being the
field at which one flux quantum penetrates the conductor).

What happens to weak localization if the normal-metal wire is attached at one
end to a superconductor? This problem has been the subject of active
research.\cite{B1992,Marmorkos,Takane,B1994,MacedoC1994,Lenssen,Nazarov} The
term $G_0$ of order $N$ is unaffected by the presence of the
superconductor.\cite{B1992} The ${\cal O}(N^0)$ correction $\delta G$ is
increased but remains universal,\cite{B1994,MacedoC1994}
\begin{equation}
  \delta G = - (2 - 8\pi^{-2})\, e^2/h \approx -1.19\, e^2/h \label{deltaG}.
\end{equation}
In all previous analytical work zero magnetic field was assumed. It was
surmised, either implicitly or explicitly,\cite{Marmorkos} that $\delta G = 0$
in the absence of ${\cal T}$\ --- but this was never actually calculated
analytically. We have now succeeded in doing this calculation and would like to
report the result, which was entirely unexpected.

We find that a magnetic field by itself is not sufficient to suppress the
weak-localization correction, but only reduces $\delta G$ by about a factor of
two. To achieve $\delta G = 0$ requires in addition the application of a
sufficiently large voltage $V$ to break the degeneracy in energy between the
electrons (at energy $eV$ above the Fermi energy $E_F$) and the
Andreev-reflected holes (at energy $eV$ below $E_F$). The electron-hole
degeneracy (${\cal D}$) is effectively broken when $eV$ exceeds the Thouless
energy $E_c = \hbar v_F \ell/L^2$ (with $v_F$ the Fermi velocity). Weak
localization coexists with a magnetic field as long as $eV \ll E_c$. Our
analytical results are summarized in Table \ref{tab1}. These results disagree
with the conclusions drawn in Ref.\ \ref{Marmorkos} on the basis of numerical
simulations. We have found that the numerical data on the weak-localization
effect was misinterpreted due to the presence of a magnetic-field dependent
contact resistance, which was not understood at that time.

\begin{center}
\vspace{0.2cm}
\begin{tabular}{|c|c|c|} \hline
 $-\delta G\, [e^2/h]$ \vphantom{{ \large $M^M_M$}} & ${\cal T}$          & \
no ${\cal T}$ \  \\ \hline \hline
 ${\cal D}$               & \ $2 - 8/\pi^2$ \vphantom{{ \large $M^M_M$}} &
$2/3$ \\ \hline
 no ${\cal D}$            & $4/3$ \vphantom{{ \large $M^M_M$}} & $0$ \\ \hline
\end{tabular}
\vspace{0.2cm}
\refstepcounter{table}
\label{tab1}
\end{center}
%
Table \ref{tab1}: Dependence of the weak-localization correction $\delta G$ of
a normal-metal wire attached to a superconductor on the presence or absence of
time-reversal symmetry (${\cal T}$) and electron-hole degeneracy (${\cal D}$).
The entry in the upper left corner was computed in Refs.\ \ref{B1994},
\ref{MacedoC1994}. \vspace{0.5cm}

Starting point of our calculation is the general relation between the
differential conductance $G = dI/dV$ of the normal-metal--superconductor (NS)
junction and the transmission and reflection matrices of the normal
region,\cite{B1992}
\begin{mathletters} \label{GNStransrefl}
\begin{eqnarray}
  && G = (4e^2/h)\, \mbox{tr}\, m(eV) m^{\dagger}(eV), \\
  && m(\varepsilon) = t'(\varepsilon)\left[1 - \alpha(\varepsilon)
r^{*}(-\varepsilon) r(\varepsilon)\right]^{-1} t^{*}(-\varepsilon),
\label{GNStransreflb}
\end{eqnarray}
where $\alpha(\varepsilon) \equiv \exp[-2{\rm i} \arccos(\varepsilon/\Delta)]$.
Eq.\ (\ref{GNStransrefl}) holds for subgap voltages $V \le \Delta/e$, and
requires also $\Delta \ll E_F$ ($\Delta$ being the excitation gap in S). We
assume that the length $L$ of the disordered normal region is much greater than
the superconducting coherence length $\xi = (\hbar v_F \ell/\Delta)^{1/2}$.
This implies that the Thouless energy $E_c \ll \Delta$. In the voltage range $V
\lesssim E_c/e$ we may therefore assume that $eV \ll \Delta$, hence $\alpha =
-1$. The $N \times N$ transmission and reflection matrices $t$, $t'$, $r$, and
$r'$ form the scattering matrix $S(\epsilon)$ of the disordered normal region
($N$ being the number of propagating modes at the Fermi level, which
corresponds to $\varepsilon = 0$). It is convenient to use the polar
decomposition
$$
  \left(\begin{array}{ll} r' & t' \\ t & r \end{array} \right) =
  \left(\begin{array}{ll} v_1 & 0 \\
                          0 & w_1 \end{array} \right)
  \left(\begin{array}{ll} {\rm i} \sqrt{{R}} &
                          \sqrt{{T}} \\
                          \sqrt{{T}} &
                          {\rm i} \sqrt{{R}} \end{array} \right)
  \left(\begin{array}{ll} v_2 & 0 \\
                          0 & w_2 \end{array} \right).
      \label{polar}
$$
Here $v_1$, $v_2$, $w_1$, and $w_2$ are $N \times N$ unitary matrices, ${T}$ is
a diagonal matrix with the $N$ transmission eigenvalues ${T}_i \in [0,1]$ on
the diagonal, and ${R} = 1 - {T}$. Using this decomposition, and substituting
$\alpha = -1$, Eq.\ (\ref{GNStransreflb}) can be replaced by

\begin{eqnarray}
 m(\varepsilon) = \sqrt{{T}(\varepsilon)} \left[1 + u(\varepsilon)
\sqrt{{R}(-\varepsilon)} u^*(-\varepsilon) \sqrt{{R}(\varepsilon)}\right]^{-1}
\nonumber \\ \ \cdot\,
u(\varepsilon) \sqrt{{T}(-\varepsilon)},\ \ u(\varepsilon) \equiv
w_2^{\vphantom{*}}(\varepsilon) w_1^*(-\varepsilon). \label{GNStransreflc}
\end{eqnarray}
\end{mathletters}%

We perform our calculations in the general framework of random-matrix theory.
The only assumption about the distribution of the scattering matrix that we
make, is that it is isotropic, i.e. that it depends only on the transmission
eigenvalues.\cite{StoneReview} In the presence of ${\cal T}$ (for $B \ll B_c$),
$S = S^{\rm T}$, hence $w_1^{\vphantom T} = w_2^{\rm T}$. (The superscript
${\rm T}$ denotes the transpose of a matrix.) If ${\cal T}$\ is broken, $w_1$
and $w_2$ are independent. In the presence of ${\cal D}$\ (for $eV \ll E_c$),
the difference between $S(eV)$ and $S(-eV)$ may be neglected. If ${\cal D}$\ is
broken, $S(eV)$ and $S(-eV)$ are independent.\cite{ergodicity} Of the four
entries in Table \ref{tab1}, the case that both ${\cal T}$\ and ${\cal D}$\ are
present is the easiest, because then $u=1$ and Eq.\ (\ref{GNStransrefl})
simplifies to\cite{B1992}
\begin{equation}
  G = (4e^2/h) {\textstyle \sum_{n}} T_n^2 (2 - T_n)^{-2}.
\end{equation}
The conductance is of the form $G = \sum_{n} f(T_n)$, known as a linear
statistic on the transmission eigenvalues. General
formulas\cite{B1994,MacedoC1994} for the weak-localization
correction to the average of a linear statistic lead directly to Eq.\
(\ref{deltaG}). The three other entries in Table \ref{tab1}, where either
${\cal T}$\ or ${\cal D}$\ (or both) are broken, are more difficult because $G$
is no longer a linear statistic. We consider these three cases in separate
paragraphs.

{(1) ${\cal D}$, no ${\cal T}$ ---} Because of the isotropy assumption, $w_1$
and $w_2$, and hence $u$, are uniformly distributed in the unitary group ${\cal
U}(N)$. We may perform the average $\langle \cdots \rangle$ over the ensemble
of scattering matrices in two steps: $\langle \cdots \rangle = \langle \langle
\cdots \rangle_{u} \rangle_T$, where $\langle\cdots\rangle_{u}$ and
$\langle\cdots\rangle_T$ are, respectively, the average over the unitary matrix
${u}$ and over the transmission eigenvalues ${T}_i$. We compute $\langle \cdots
\rangle_u$ by an expansion in powers of $N^{-1}$. To integrate the rational
function (\ref{GNStransrefl}) of $u$ over ${\cal U}(N)$, we first expand it
into a geometric series and then use the general rules for the integration of
polynomials of $u$.\cite{Creutz,Mello,BrouwerBeenakker} The polynomials we need
are
\begin{mathletters}
\begin{eqnarray}
  && \langle G \rangle_u = {4e^2 \over h}\, \sum_{p,q=0}^{\infty} M_{pq}, \\
  && M_{pq} = \langle \mbox{tr}\, {T} (u \sqrt{{R}} u^* \sqrt{{R}})^p u {T}
u^{\dagger} (\sqrt{{R}} u^{\rm T} \sqrt{{R}} u^{\dagger})^q \rangle_{u}.
\end{eqnarray}
\end{mathletters}%
Neglecting terms of order $N^{-1}$, we find
\begin{eqnarray*}
   && M_{pq} = \left\{ \begin{array}{l} N {\tau}_1^2 (1 -
{\tau}_1^{\vphantom{2}})^{2p} \ \mbox{if $p = q$,} \\
 {\tau}_1({\tau}_1^2 + {\tau}_1^{\vphantom{2}} - 2{\tau}_2^{\vphantom{2}}) (1 -
{\tau}_1^{\vphantom{2}})^{p + q - 1} - 
 2\, \mbox{min}(p,q) \\ \ \times\, {\tau}_1^2 ({\tau}_1^2 -
{\tau}_2^{\vphantom{2}}) (1 - {\tau}_1^{\vphantom{2}})^{p+q-2} \ \mbox{if $|p -
q|$ odd,} \\ 0 \ \mbox{else}, \end{array} \right.
\end{eqnarray*}
where we have defined the moment ${\tau}_k = N^{-1} \sum_i {T}_i^k$. The
summation over $p$ and $q$ leads to
\begin{equation}
 {h \over 4e^2}\, \langle G \rangle_{u} = {N {\tau}_1 \over 2 - {\tau}_1} - {4
{\tau}_1 - 2 {\tau}_1^2 + 2 {\tau}_1^3 - 4 {\tau}_2 \over {\tau}_1 (2 -
{\tau}_1)^3}. \label{GNSavgu}
\end{equation}
It remains to average over the transmission eigenvalues. Since ${\tau}_k$ is a
linear statistic, we know that its sample-to-sample fluctuations $\delta
{\tau}_k \equiv {\tau}_k - \langle {\tau}_k \rangle$ are an order $1/N$ smaller
than the average.\cite{StoneReview} Hence
\begin{equation}
  \langle f({\tau}_k) \rangle_T = f(\langle {\tau}_k \rangle) [1 + {\cal
O}(N^{-2})], \label{Nofluct}
\end{equation}
which implies that we may replace the average of the rational function
(\ref{GNSavgu}) of the ${\tau}_k$'s by the rational function of the average
$\langle {\tau}_k \rangle$. This average has the $1/N$ expansion
\begin{equation}
  \langle {\tau}_k \rangle = \langle {\tau}_k \rangle_0 + {\cal O}(N^{-2}),
\label{TauAvgExpand}
\end{equation}
where $\langle {\tau}_k \rangle_0$ is ${\cal O}(N^0)$. There is no term of
order $N^{-1}$ in the absence of ${\cal T}$. From Eqs.\
(\ref{GNSavgu})--(\ref{TauAvgExpand}) we obtain the $1/N$ expansion of the
average conductance,
\begin{equation}
 {h \over 4e^2}\, \langle G \rangle = {N\langle {\tau}_1 \rangle_0 \over 2 -
\langle {\tau}_1 \rangle_0} - {4 \langle {\tau}_1 \rangle_0 - 2 \langle
{\tau}_1 \rangle_0^2 + 2\langle {\tau}_1 \rangle_0^3 - 4 \langle {\tau}_2
\rangle_0 \over \langle {\tau}_1 \rangle_0(2 - \langle {\tau}_1 \rangle_0)^3} +
{\cal O}(N^{-1}). \label{GNSavgbeta2}
\end{equation}
Eq.\ (\ref{GNSavgbeta2}) is generally valid for any isotropic distribution of
the scattering matrix. We apply it to the case of a disordered wire in the
limit $N \rightarrow \infty$, $\ell/L \rightarrow 0$ at constant $N\ell/L$. The
moments $\langle {\tau}_k \rangle_0$ are given by\cite{MelloStone}
\begin{equation}
  \langle {\tau}_1 \rangle_0 = \ell/L,\  \langle {\tau}_2 \rangle_0 =
\case{2}{3}\ell/L. \label{DisorderedWire}
\end{equation}
Substitution into Eq.\ (\ref{GNSavgbeta2}) yields the weak-localization
correction $\delta G = -\case{2}{3} e^2/h$, cf.\ Table \ref{tab1}.

{(2) ${\cal T}$, no ${\cal D}$ ---} In this case one has $u^{\dagger}(-eV) =
u(eV)$ and $u(eV)$ is uniformly distributed in ${\cal U}(N)$. A calculation
similar to that in the previous paragraph yields for the average over $u$:
\begin{eqnarray}
  {h \over 4e^2}\, \langle G \rangle_{u} &=& N {\tau}_{1+}^{\vphantom{2}}
{\tau}_{1-}^{\vphantom{2}}({\tau}_{1+}^{\vphantom{2}} +
{\tau}_{1-}^{\vphantom{2}} - {\tau}_{1+}^{\vphantom{2}}
{\tau}_{1-}^{\vphantom{2}})^{-1} + ({\tau}_{1+}^{\vphantom{2}} +
{\tau}_{1-}^{\vphantom{2}} - {\tau}_{1+}^{\vphantom{2}}
{\tau}_{1-}^{\vphantom{2}})^{-3} \nonumber \\ && \ \times\, \left[2
{\tau}_{1+}^2 {\tau}_{1-}^2 - {\tau}_{1+}^3 {\tau}_{1-}^2 - {\tau}_{1+}^2
{\tau}_{1-}^3 - {\tau}_{2+}^{\vphantom{2}} {\tau}_{1-}^2 - {\tau}_{1+}^2
{\tau}_{2-}^{\vphantom{2}} + {\tau}_{2+}^{\vphantom{2}} {\tau}_{1-}^3 +
{\tau}_{1+}^3 {\tau}_{2-}^{\vphantom{2}} \right],
\end{eqnarray}
where we have abbreviated ${\tau}_{k\pm} = {\tau}_k(\pm eV)$. The next step is
the average over the transmission eigenvalues. We may still use Eq.\
(\ref{Nofluct}), and we note that $\langle {\tau}_k(\varepsilon) \rangle \equiv
\langle {\tau}_k \rangle$ is independent of $\varepsilon$. (The energy scale
for variations in $\langle {\tau}_k(\varepsilon) \rangle$ is $E_F$, which is
much greater than the energy scale of interest $E_c$.) Instead of Eq.\
(\ref{TauAvgExpand}) we now have the $1/N$ expansion
\begin{equation}
  \langle {\tau}_k \rangle = \langle {\tau}_k \rangle_0 + N^{-1} \delta
{\tau}_k + {\cal O}(N^{-2}),
\end{equation}
which contains also a term of order $N^{-1}$ because of the presence of ${\cal
T}$. The $1/N$ expansion of $\langle G \rangle$ becomes
\begin{equation}
  {h \over 4e^2}\, \langle G \rangle = {N \langle {\tau}_1 \rangle_0 \over 2 -
\langle {\tau}_1 \rangle_0} + {2 \delta {\tau}_1 \over (2 - \langle {\tau}_1
\rangle_0)^2} + {2\langle {\tau}_1 \rangle_0^2- 2\langle {\tau}_1 \rangle_0^3 -
2\langle {\tau}_2 \rangle_0 + 2\langle {\tau}_1 \rangle_0 \langle {\tau}_2
\rangle_0 \over \langle {\tau}_1 \rangle_0(2 - \langle {\tau}_1 \rangle_0)^3}
\label{GNSavgbeta1V} + {\cal O}(N^{-1}).
\end{equation}
For the application to a disordered wire we use again Eq.\
(\ref{DisorderedWire}) for the moments $\langle {\tau}_k \rangle_0$, which do
not depend on whether ${\cal T}$\ is broken or not. We also need $\delta
{\tau}_1$, which in the presence of ${\cal T}$\ is given by\cite{MelloStone}
$\delta {\tau}_1 = -\case{1}{3}$. Substitution into Eq.\ (\ref{GNSavgbeta1V})
yields $\delta G = -\case{4}{3} e^2/h$, cf.\ Table \ref{tab1}.

{(3) no ${\cal T}$, no ${\cal D}$ ---} Now $u(eV)$ and $u(-eV)$ are
independent, each with a uniform distribution in ${\cal U}(N)$. Carrying out
the two averages over ${\cal U}(N)$ we find
\begin{equation}
 {h \over 4e^2}\, \langle G \rangle_{u} = {N {\tau}_{1+}^{\vphantom{2}}
{\tau}_{1-}^{\vphantom{2}} \over {\tau}_{1+}^{\vphantom{2}} +
{\tau}_{1-}^{\vphantom{2}} - {\tau}_{1+}^{\vphantom{2}}
{\tau}_{1-}^{\vphantom{2}}}. \label{GNSavgub2V}
\end{equation}
The average over the transmission eigenvalues becomes
\begin{equation}
  {h \over 4e^2}\, \langle G \rangle = {N \langle {\tau}_1 \rangle_0 \over 2 -
\langle {\tau}_1 \rangle_0} + {\cal O}(N^{-1}), \label{GNSavgbeta2V}
\end{equation}
where we have used that $\delta {\tau}_1 = 0$ because of the absence of ${\cal
T}$. We conclude that $\delta G = 0$ in this case, as indicated in Table
\ref{tab1}.

This completes the calculation of the weak-localization correction to the
average conductance. Our results, summarized in Table \ref{tab1}, imply a
universal $B$ and $V$-dependence of the conductance of an NS microbridge.
Raising first $B$ and then $V$ leads to two subsequent increases of the
conductance, while raising first $V$ and then $B$ leads first to a decrease and
then to an increase.

So far we have only considered the ${\cal O}(N^0)$ correction $\delta G$ to
$\langle G \rangle = G_0 + \delta G$. What about the ${\cal O}(N)$ term $G_0$?
{}From Eqs.\ (\ref{GNSavgbeta2}), (\ref{GNSavgbeta1V}), and
(\ref{GNSavgbeta2V}) we see that if either ${\cal T}$\ or ${\cal D}$\ (or both)
are broken,
\begin{equation}
  G_0 = {4e^2 \over h}\, {N \langle {\tau}_1 \rangle_0 \over 2 - \langle
{\tau}_1 \rangle_0} = (2e^2 / h)\, N \left( \case{1}{2} + {L / \ell}
\right)^{-1}. \label{contact1}
\end{equation}
In the second equality we substituted\cite{MelloStone} $\langle {\tau}_1
\rangle_0 = (1 + L/\ell)^{-1}$, which in the limit $\ell/L \rightarrow 0$
reduces to Eq.\ (\ref{DisorderedWire}). If both ${\cal T}$\ and ${\cal D}$\ are
unbroken, then we have instead the result\cite{BeenakkerRejaeiMelsen}
\begin{equation}
  G_0 = (2e^2 /h)\, N \left[ 1 + {L/\ell} + {\cal O}(\ell/L) \right]^{-1}.
\label{contact2}
\end{equation}
The difference between Eqs.\ (\ref{contact1}) and (\ref{contact2}) is a contact
resistance, which equals $h/4Ne^2$ in Eq.\ (\ref{contact1}) but is twice as
large in Eq.\ (\ref{contact2}). In contrast, in a normal-metal wire the contact
resistance is $h/2Ne^2$, independent of $B$ or $V$. The $B$ and $V$-dependent
contact resistance in an NS junction is superimposed on the $B$ and
$V$-dependent weak-localization correction. Since the contribution to $\langle
G \rangle$ from the contact resistance is of order $(e^2/h) N (\ell/L)^2$,
while the weak-localization correction is of order $e^2/h$, the former can only
be ignored if $N (\ell/L)^2 \ll 1$. This is an effective restriction to the
diffusive metallic regime, where $\ell/L \ll 1$ and $N\ell/L \gg 1$. To measure
the weak-localization effect without contamination from the contact resistance
if $N(\ell/L)^2$ is not $\ll 1$, one has two options: (1) Measure the
$B$-dependence at fixed $V \gg E_c/e$; (2) Measure the $V$-dependence at fixed
$B \gg B_c$. In both cases we predict an increase of the conductance, by an
amount $\case{4}{3} e^2/h$ and $\case{2}{3}e^2/h$, respectively. In contrast,
in the normal state weak localization leads to a $B$-dependence, but not to a
$V$-dependence.

We performed numerical simulations similar to those of Ref.\ \ref{Marmorkos} in
order to test the analytical predictions. The disordered normal region was
modeled by a tight-binding Hamiltonian on a square lattice (lattice constant
$a$), with a random impurity potential at each site (uniformly distributed
between $\pm \frac{1}{2} U_d$). The Fermi energy was chosen at $E_F = 1.57 U_0$
from the band bottom ($U_0 = \hbar^2/2ma^2$). The length $L$ and width $W$ of
the disordered region are $L = 167 a$, $W = 35 a$, corresponding to $N = 15$
propagating modes at $E_F$. The mean free path is obtained from the conductance
$G = (2e^2/h) N (1 + L/\ell)^{-1}$ of the normal region in the absence of
${\cal T}$. The scattering matrix of the normal region was computed numerically
at $\varepsilon = \pm eV$, and then substituted into Eq.\ (\ref{GNStransrefl})
to obtain the differential conductance.

In Fig.\ \ref{fig1} we show the $V$-dependence of $G$ (averaged over some
$10^3$ impurity configurations), for three values of $\ell$. The left panel is
for $B=0$, the right panel for a flux of $6\, h/e$ through the disordered
region. The $V$-dependence for $B=0$ is mainly due to the contact resistance
effect of order $N(\ell/L)^2$, and indeed one sees that the amount by which $G$
increases depends significantly on $\ell$.\cite{AnalCalc} The $V$-dependence in
a ${\cal T}$-violating magnetic field is entirely due to the weak-localization
effect, which should be insensitive to $\ell$ (as long as $\ell/L \ll 1 \ll
N\ell/L$). This is indeed observed in the simulation. Quantitatively, we would
expect that application of a voltage increases $\langle G \rangle$ by an amount
$\case{2}{3}e^2/h$ for the three curves in the right panel, which agrees very
well with what is observed. In the absence of a magnetic field the analytical
calculation predicts a net increase in $\langle G \rangle$ by $0.79$, $0.46$,
and $0.25\, \times \, e^2/h$ (from top to bottom), which is again in good
agreement with the simulation.

In conclusion, we have shown that weak localization can coexist with a
time-reversal symmetry breaking magnetic field in a disordered
normal-metal--superconductor junction. One needs to apply a magnetic field and
to raise the voltage in order to suppress the weak-localization correction to
the conductance. This leads to an unusual $B$ and $V$-dependence of the
differential conductance, which should be observable in experiments.

We thank M.\ J.\ M.\ de Jong and J.\ A.\ Melsen for help with the numerical
simulations. This work was supported by the Dutch Science Foundation NWO/FOM.


\begin{figure}[h]
\caption{\label{fig1} Numerical simulation of the voltage dependence of the
average differential conductance for $B=0$ (left panel) and for a flux $6\,
h/e$ through the disordered normal region (right panel). The filled circles are
for an NS junction; the open circles represent the $V$-independent conductance
in the normal state. The three sets of data points correspond, from top to
bottom, to $\ell/L = 0.31$, $0.23$, and $0.18$, respectively. The arrows
indicate the theoretically predicted net increase of $\langle G \rangle$
between $V=0$ and $V \gg E_c/e$.}

\end{figure}


\begin{references}
\bibitem{SO} Spin-orbit scattering is assumed to be negligible, otherwise a
positive quantum correction appears.
\bibitem{WIRE} The restriction to a wire geometry is essential for the
universality. In a square or cube geometry, the weak-localization correction is
$L$ and $\ell$-dependent.
\bibitem{AnalCalc} Our analytical calculation refers to the net increase of
$\langle G \rangle$ between $V=0$ and $V \gg E_c/e$. We can not describe the
non-monotonic $V$-dependence at intermediate $V$, observed in the simulation
for $B=0$.
\end{references}
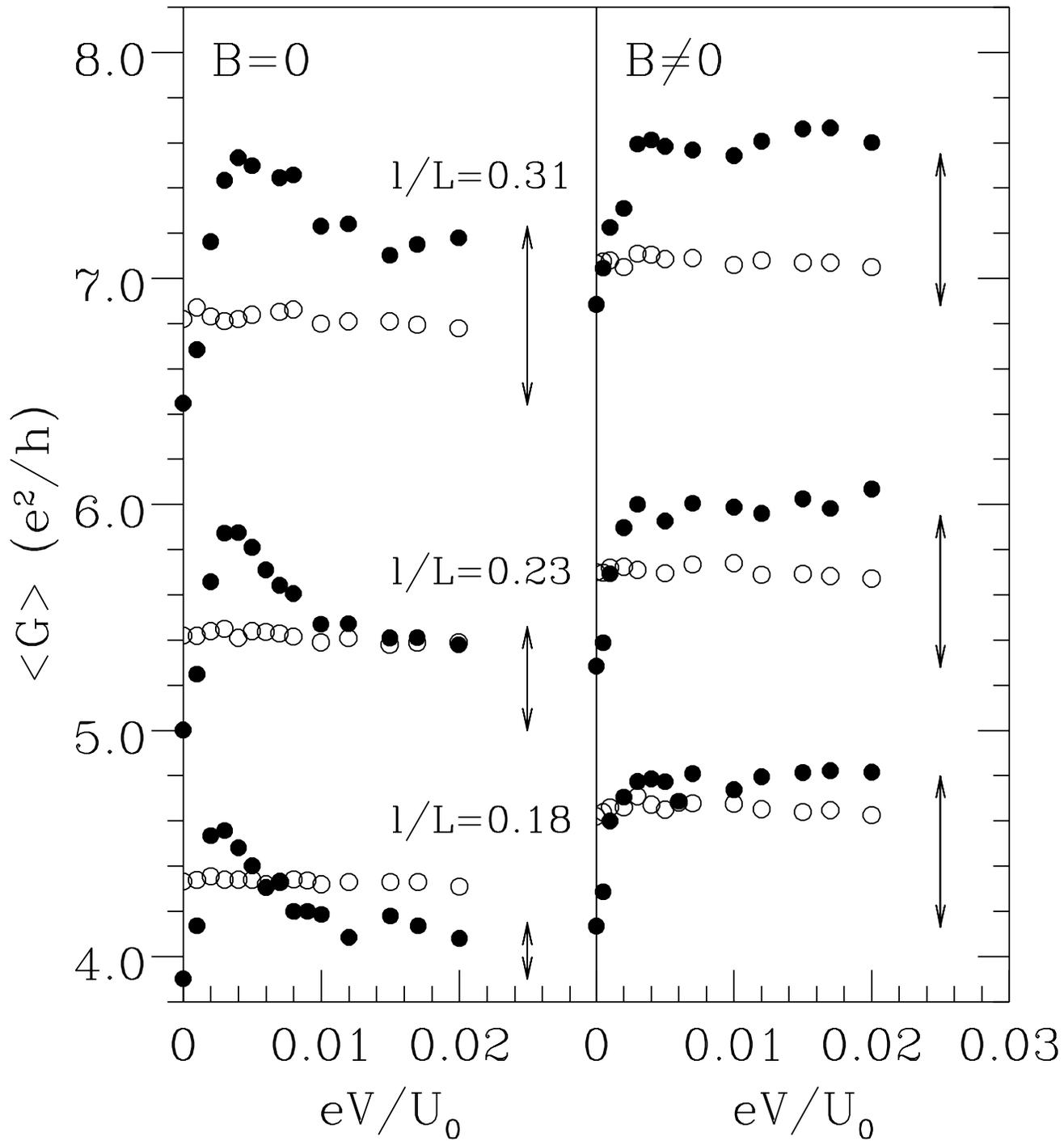
\end{document}